\documentstyle[12pt,psfig,twoside]{article} 

\begin{document}
\hbadness=10000
\pagenumbering {arabic}
\pagestyle{myheadings}
\markboth{J. Letessier, J. Rafelski and A. Tounsi}
{Gluon production, cooling and entropy in nuclear collisions}
 
\title{ GLUON PRODUCTION, COOLING AND ENTROPY IN NUCLEAR COLLISIONS}
\author{$\ $\\Jean LETESSIER$^{1}$,\ \  Johann RAFELSKI$^{1,2[\dagger]}$,\ \   
and Ahmed TOUNSI$^{1}$\\
$\ $\\
{$^{1}$}Laboratoire de Physique Th\'eorique et Hautes Energies,
Paris\thanks{Unit\'e  associ\'ee au CNRS, Universit\'e PARIS 7, Tour 24,
5\`e \'et., 2 Place Jussieu, F-75251 CEDEX 05.}\\ 
{$^{2}$}Department of Physics, University of Arizona, Tucson, AZ 85721, USA.\\}

\date{} 

\maketitle

\begin{abstract}
 
\noindent
We study the cooling (heating) of a glue-parton gas due to production 
(destruction) of particles and determine the associated production of
entropy. We incorporate sharing of the system energy among a changing
number of particles. We find that the entropy of an evolving 
glue-parton gas changes in an insignificant range once the initial
high temperature state has been formed, despite a great change in
particle number and temperature.


\end{abstract}
\begin{center}
Published in {\it Phys. Rev.} C 50 (1994) 406--409.
\end{center}
\vfill
{\bf PAR/LPTHE/93--51}\hfill{\bf October 1993}
\eject

In nuclear collisions carried out at energies per nucleon considerably
greater than the nucleon rest mass we observe a very high final state
particle multiplicity, with which is associated an unusually large
entropy content \cite{Let93}. There are two
questions which immediately come to mind when looking at this result:
i) when and how is entropy produced in a quantum process, such as is a
nuclear collision, and ii) how is the particle production related to
entropy production?
 
We will address here the second question and investigate in a
quantitative manner the increase in entropy during the evolution of
the primordial gluon-parton system \cite{Shu92,Gei92b,Bir93}. In our
approach, as in every prior work up to date, we do not view the
nuclear collision reaction as a reversible quantum process, but rather
as a series of disorder increasing independent reactions. Thus the
measurement of (conventionally defined) entropy can be performed at
any instant during the collisions \cite{Gei92b}, which approach may be
seen in contradiction to the prevailing understanding of reversibility
of quantum evolution. This is a pragmatic approach in view of the
final state formed in the collisions, which is displaying high
particle multiplicity. The view we and other workers take is that a
solution to the more fundamental question i) will justify the
pragmatic approach taken today regarding question ii), where we
proceed as if the nuclear collision reaction was a classical process
involving particle production.
 
The physical systems of interest to us here are characterized by
relatively large elastic and quasi elastic cross sections, which are
typically 10--30 times larger than the particle number changing
cross sections. Therefore it can be assumed that the so called kinetic
equilibrium characterized by the sharing of the available energy
between the particles, and the establishment of a common
temperature, is reached instantaneously as compared to the slowly
evolving particle number. We call a particle species to be in {\it
Hagedorn}  equilibrium \cite{Hag73}, when all phase space cells are
occupied according to a distribution which maximize the entropy
content at some given energy. It is interesting to note here that
there is so far apparently no
general proof of Boltzmann's H-theorem to be found in literature for
relativistic systems with particle number non-conserving reactions
\cite{deG80}. As expected we find below that Hagedorn
equilibrium corresponds to the standard Bose/Fermi quantum
distribution of particles, which give a maximum of system entropy at
fixed energy. This is a preliminary
step towards such a proof, which requires further that the particle
numbers, to be considered in kinetic theory, evolve in time towards
the Hagedorn equilibrium --- while for a select system
\cite{OH93} this has been argued, a general discussion of this issue
remains an open problem. In our detailed discussion we focus on
the situation found in studies of the `hot glue' initial state
proposed by Shuryak \cite{Shu92}. We will show that despite a great
increase in particle number as the system approaches Hagedorn
equilibrium relatively little additional entropy is generated.
 
We begin with the considerations presented by Landau and
Lifshitz \cite{Lan?}. The usual definition of entropy
for each particle species denoted by index $l$ is:
\begin{equation}\label{entro1}
S_{\rm B,F}^l=\int\!d\omega\,\left[\pm(1\pm n_l)\ln(1\pm n_l)
-               n_l\ln n_l\right]\, ,
\end{equation}
where the integral is over the conventional phase space
$d\omega=d^3\!p d^3\!x/(2\pi\hbar)^3$ (we henceforth choose units such
that $\hbar=c=1$). The upper sign (here $+$) applies to the Bose
particles (`B'), while the lower sign (here $-$) applies to the Fermi
particles (`F'). The Boltzmann limit follows when the
occupation $n_l$ of the phase space by one of the particle species is
small compared to unity. 
Given a number of particles of each species, $N^l\equiv\int\!d\omega\,
n_l$, and the energy of the system $E=\sum_l E^l\equiv\int\!d\omega\,
\epsilon_l n_l$, where $\epsilon_l=\sqrt{m_l^2+p^2}$, 
the form of $n_l(p,x)$ is obtained by maximizing the quantity ${\cal
T}^l[n_l]={\cal S}^l-(\alpha_l+\beta\epsilon_l)n_l$
as a functional of $n_l$. $\alpha^l$ permits to choose any given
number of particles of kind $l$ and $\beta=1/T$, which permits to keep
the energy constant, is identified as the inverse temperature.
Reinserting the Bose `B' and Fermi `F' quantum distribution functions
\begin{equation}\label{qdist}
n_{\rm B,F}=
     {1\over{ {\rm e}^{\beta\epsilon+\alpha}\mp 1}}
\end{equation}
which follow from this consideration into the definition of entropy,
Eq.\,(\ref{entro1}) we obtain the explicit form \cite{SH92}:
\begin{eqnarray}
S^l_{\rm B,F}&&\hspace{-0.6cm}=\!\int\!d\omega\, \left[
     {{\beta\epsilon_l+\alpha_l}\over 
     { {\rm e}^{\beta\epsilon_l+\alpha_l} \mp 1}
     } \mp\ln(1\mp {\rm e}^{-\beta\epsilon_l-\alpha_l}) \right]
     \, .\label{entro2}
\end{eqnarray}
The effect of a changing particle number appears in two different ways:
as a factor 
\begin{equation}
\gamma_l\equiv {\rm e}^{-\alpha_l}
\end{equation}
in front of all Boltzmann exponential present in above equations and
secondly, as a coefficient of an additional additive, particle number
proportional term. We can thus obtain $S_l$, Eq.\,(\ref{entro2}), as a
derivative of the partition function $\ln{{\cal Z}^l}(\beta,\alpha_l)  
=\mp\!\int\!d\omega\ln(1\mp {\rm e}^{-\beta\epsilon_l-\alpha_l})$\,.
 
In the Boltzmann approximation the factor $\gamma_l$ becomes a
normalization factor which describes the average occupancy of the
phase space relative to the equilibrium value \cite{Raf91}, the
additive $\alpha_l$-term in Eq.\,(\ref{entro2}) contributes to the
entropy per particle change as the occupancy changes; in particular we
note that in the Boltzmann limit: 
\begin{equation}\label{SNBol}
{S^l\over N^l}=
\left.{S^l\over N^l}\right|_{\rm eq}+\ln\gamma^{-1}_l\, .
\end{equation}
When we omit the last term from the discussion of the entropy of a
system, we find that the entropy evolves along with the number of 
particles.  Considering this term we see that as the occupancy of
the particle phase space increases at fixed $\beta$, the entropy per
particle decreases. This observation is of course not in contradiction
to our expectation that the {\it total} entropy should increase as the
particle number increases. The situation appears to be, however, much
less trivial when we look at what happens as the occupancy of phase
space approaches the equilibrium value from above $1<\gamma_l\to1\,.$
For the entropy must increase while the number of particles decreases.
This can clearly only be the case if we consider how particle
annihilation heats the system at given energy $E$. Similarly, we must
consider when studying the increase of entropy during the approach
from below to the equilibrium occupancy, how the production of
particles cools the system.
 
We find the required change in $\beta$ for systems without conserved
quantum numbers from: 
\begin{equation}
dE=\sum_l{\partial E^l\over \partial \alpha_l}d\alpha_l
     +{\partial E\over \partial \beta}d\beta\, ,
\end{equation}
which is inserted into the change in entropy to yield:
\begin{equation}
\label{dS}
\hspace{-0.4cm}d{S}=
     dE{{\partial {S}/\partial \beta}\over
     {\partial E/\partial \beta}}
+     \hspace{-0.1cm}\sum_l\! d\alpha_l\! \left[
      {{\partial {S}^l}\over{\partial \alpha_l}}
-     {{\partial {S}}\over{\partial \beta}}
      {{\partial E^l/\partial \alpha_l}\over
      {\partial E/\partial\beta}}\right] .
     \end{equation}
It is our hypothesis that the system is isolated,  hence $dE=0$.
Considerable cancellation occurs when we evaluate Eq.\,(\ref{dS}):
\begin{eqnarray}
\left.dS^l\right\vert_{E}=d\gamma_l\int\!{d\omega
\over{\left(1\mp\gamma_le^{-\beta\epsilon_l}\right)^2}}
     \biggl[\alpha_l e^{-\beta\epsilon_l}
-     \epsilon_l e^{-\beta\epsilon_l}
     {{\sum_k\int\!d\bar \omega_k\, \alpha_k
     \epsilon_k \gamma_k e^{-\beta\epsilon_k}}\over
     {\sum_k\int\!d\bar \omega_k\,\epsilon_k 
     \epsilon_k \gamma_k e^{-\beta\epsilon_k}
}}\biggr]\,.
\end{eqnarray}
 
We see that indeed the change in entropy vanishes exactly when all
$\alpha_l$ vanish, that is when $\gamma_l\to 1$. Furthermore we can
see using Schwartz' inequality that for any $\gamma_l<1$
the change $dS$ is positive for positive $d\gamma_l$ while for
$\gamma_l>1$ the change is negative, proving that the full Hagedorn
equilibrium ($\gamma_l\to 1$) gives maximum of entropy at
fixed total energy.
 
We consider now a systems of gluons in which numerous different
reactions such as $G^*\to GG$, $GG\to GGG,\ \ldots,$  
are thought to increase the number of gluons rapidly and to lead
to development of Hagedorn equilibrium. Our objective is to see the
usually ignored effect of cooling due to particle production, and to
determine the magnitude of the associated entropy increase. Our
calculations are set up such that the Hagedorn equilibrium results at
$T=0.250$ GeV --- due to energy conservation at fixed volume the 
energy density is constant as function of
$\gamma_{\rm G}$ (index `G' is henceforth dropped) and takes the value
2.66 GeV/fm$^3$. When we consider a thermal mass for gluons, which we
take for illustrative purposes to be $m_{\rm G}^{\rm th}=0.200$ GeV,
we also choose to have Hagedorn equilibrium at $T=0.250$ GeV and
therefore the constant energy density drops to 1.89 GeV/fm$^3$.

\begin{figure}[t]
\begin{minipage}[t]{0.475\textwidth}
\vspace{-1.6cm}
\centerline{\hspace{0.2cm}\psfig{figure=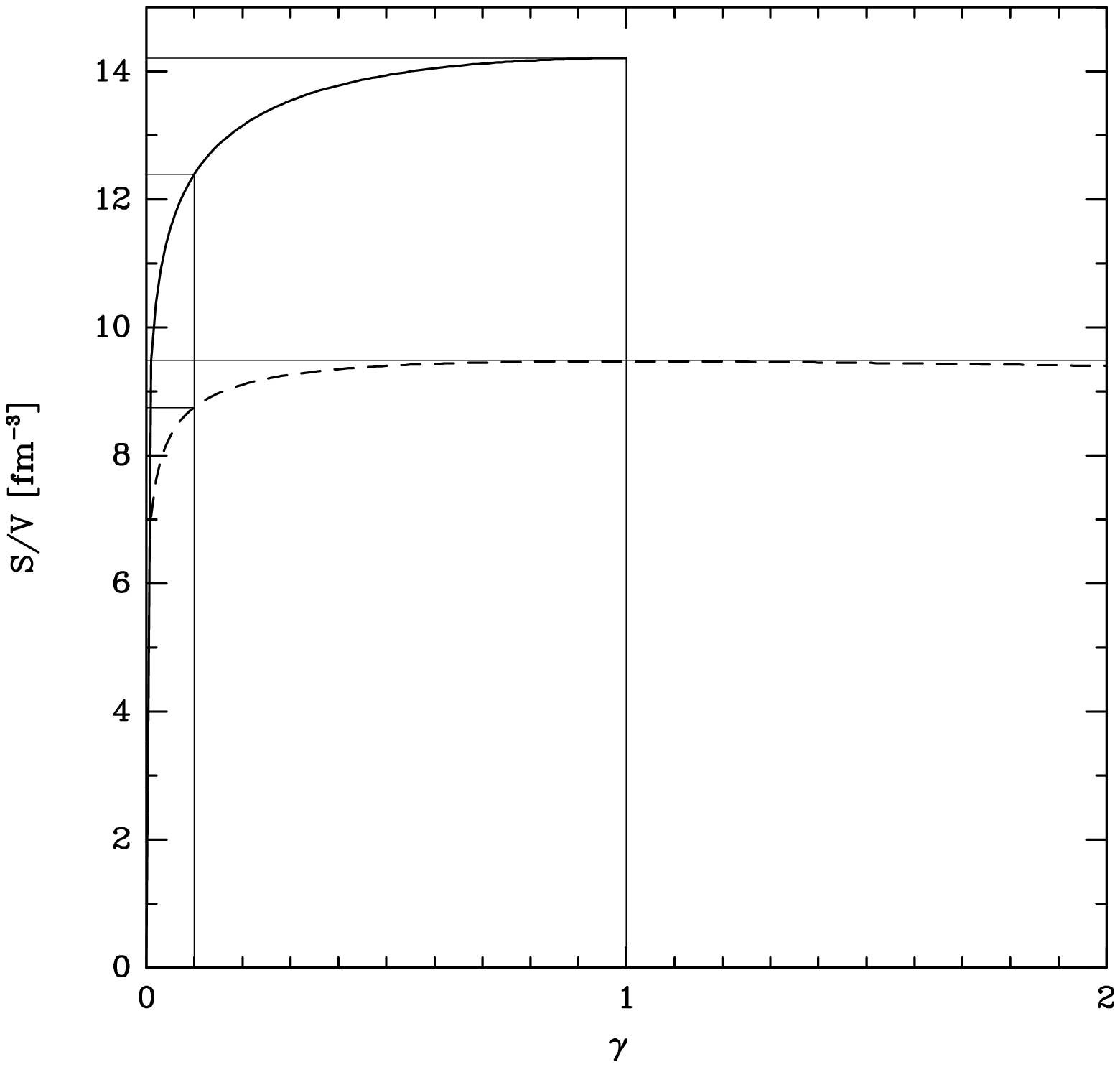,height=11.5cm}}

\vspace{ -2.cm}
\caption{\small
Entropy density ${S}/V$ (units 1/fm$^3$) at fixed energy density
$E/V=2.66$ GeV/fm$^3$ for $m_{\rm G}=0$ (solid line) and at $E/V=1.89$
GeV/fm$^3$ for $m_{\rm G}=0.200$ GeV (dashed line) for a (gluon) Bose
gas as function of the chemical occupancy $\gamma$.
\protect\label{F1}
}
\end{minipage}\hfill
\begin{minipage}[t]{0.475\textwidth}
\vspace{- 1.6cm}
\centerline{\hspace{0.2cm}\psfig{figure=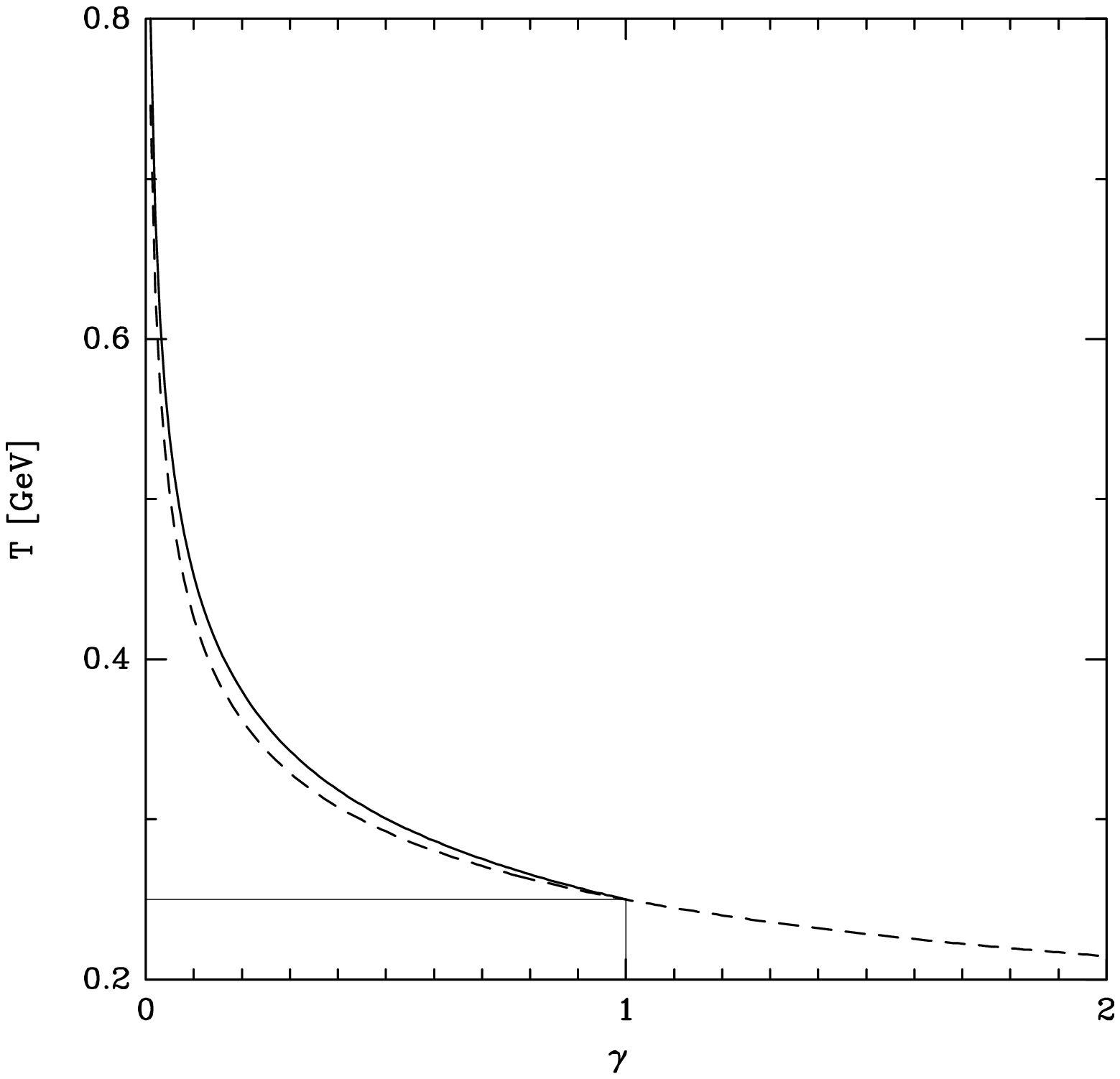,height=11.5cm}}

\vspace{ -2.cm}
\caption{\small
The temperature $T$ as function of the chemical occupancy $\gamma$.
Lines as in Fig. \protect\ref{F1}; equilibrium point $\gamma=1$ has been chosen to
occur at $T=0.250$ GeV.
\protect\label{F2}
}
\end{minipage}
\vspace{ -0.5cm}
\end{figure}

\begin{figure}[t]
\begin{minipage}[t]{0.475\textwidth}
\vspace{ -1.6cm}
\centerline{\hspace{0.2cm}\psfig{figure=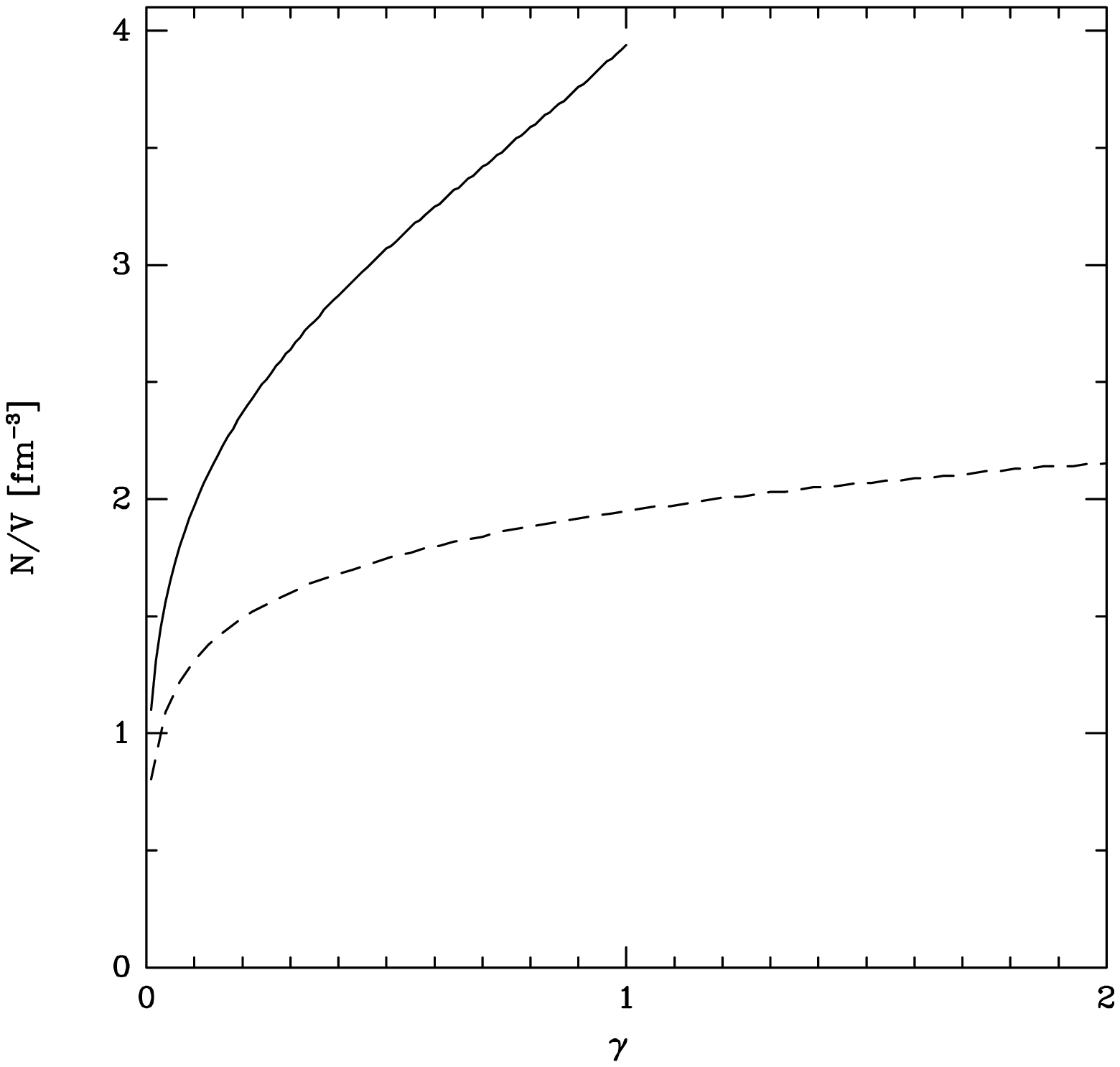,height=11.5cm}}\vspace{-2cm}

\vspace{ -0.5cm}
\caption
{\small  
Particle density $N/V$ (units 1/fm$^3$) as function of the chemical
occupancy $\gamma$. Lines as in Fig. \protect\ref{F1}.
\protect\label{F3}
}
\end{minipage}\hfill
\begin{minipage}[t]{0.475\textwidth}
\vspace{ -1.6cm}
\centerline{\hspace{0.2cm}\psfig{figure=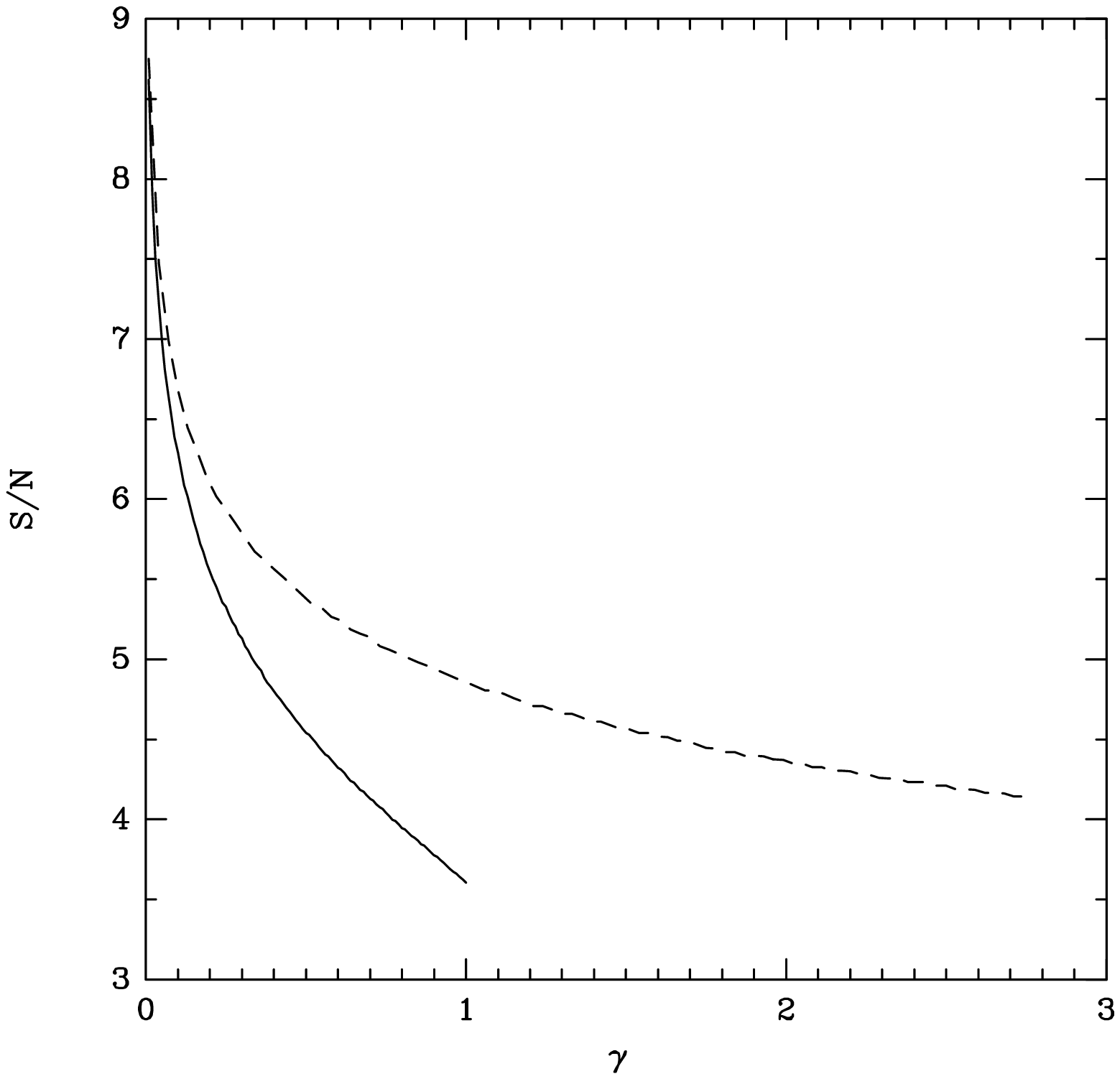,height=11.5cm}}\vspace{-2cm}

\vspace{ -0.5cm}
\caption
{\small  
Entropy per particle $S/N$ for a (gluon) Bose gas as function of the
chemical occupancy $\gamma$. Lines as in Fig. \protect\ref{F1}.
\protect\label{F4}
}
\end{minipage}
\vspace{ -0.2cm}

\end{figure}

We show in Fig.\,\ref{F1} the entropy density ${S}/V$ (units 1/fm$^3$)
as function of $\gamma$. We note that for massless gluons we can not
proceed beyond $\gamma=1$ because of Bose-condensation phenomena not
considered further here. We show by dashed lines the same calculation
carried out at $m_{\rm G}^{\rm th}=0.200$ GeV, which allows us to
continue the calculation up to $\gamma\sim2.7$. The remarkable feature
to be noted in the Fig.\,\ref{F1} is the appearance of the very weak
maximum at $\gamma=1$ --- we observe that for all practical purposes
in the hot glue picture \cite{Shu92} all entropy of the system is
already created when the particle number evolution begins at
$\gamma\sim 0.1$ \cite{Bir93}. However, the temperature $T$ shown in
Fig.\,\ref{F2} evolves rapidly in the interval
$0.1<\gamma<1$. While the total entropy increases by 15\% (respectively 11\%) 
the temperature drops from 0.45 GeV (respectively 0.43 GeV) to 0.250 GeV 
for $m_{\rm G}^{\rm th}=0.$ (respectively $m_{\rm G}^{\rm th}=0.2$ GeV), this
decrease being driven here solely by particle production (no expansion
cooling). 
 
As we can further note in Fig.\,\ref{F3} the glue number density
increases considerably in the same region --- it more than doubles for
massless gluons. In Fig.\,\ref{F3} we also note that the glue number
density increases monotonically as we pass $\gamma=1$ for $m_{\rm
G}>0$. On the other hand, Fig.\,\ref{F4} shows that the entropy per
particle drops continuously as could be expected from the qualitative
result obtained in the Boltzmann limit, Eq.\,(\ref{SNBol}).
 
We note that in the expanding glue fireball there are no other known
sources of entropy, in particular since the expansion is believed to
be a largely entropy conserving hydrodynamical flow process. We will
now address in qualitative terms the interplay between expansion and
particle production cooling. In principle we need to know with some
precision the relative rates of change of the volume $V$ and the
occupancy factor $\gamma$. Details of the such calculations are by
necessity model dependent, but we can obtain some interesting
qualitative insights pursuing some general relations. For (nearly)
massless gas we have:
\begin{eqnarray}
S&&\hspace{-0.6cm}={4\over 3} c_1 V T^3 \gamma + c_2 V T^3 \gamma\alpha\, ,
\nonumber\\
E_{\rm th}&&\hspace{-0.6cm}=E_0-E_{\rm flow}=c_1 V T^4 \gamma\label{flow1}\, .
\end{eqnarray}
Both $c_1(\gamma,\beta m)$, and $c_2(\gamma,\beta m)$ are well known
factors and can be exactly computed. For $\beta m\to 0$ we find:\
\begin{equation}
\frac{c_1}{c_2}=3\ \frac{\displaystyle\sum^\infty_{n=1}
\frac{\gamma^{n-1}}{n^4}}
{\displaystyle\sum^\infty_{n=1}\frac{\gamma^{n-1}}{n^3}}=\left\{
\begin{array}{rl}
2.7 \ {\rm for}\ \gamma&\hspace{-0.3cm}=1\ ,\\\ \\
3 \ {\rm for}\ \gamma&\hspace{-0.3cm}=0\ .
\end{array}\right.
\end{equation}
In the Boltzmann limit the factors $c_1,c_2$ are independent of
$\alpha=\ln\gamma^{-1}$. From Eq.\,(\ref{flow1}) we obtain using
$c_1/c_2\simeq 3$:
\begin{eqnarray}
S\simeq S_0{T_0\over T}{{1+\alpha/4}\over{1+\alpha_0/4}}
     \left(1-{E_{\rm flow}\over E_0}\right)\, .\label{flow2}
\end{eqnarray}
When $E_{\rm flow}=0$ this is a condition satisfied by the results we
presented before. Because of the appearance of the flow factor in
Eq.\,(\ref{flow2}) which reduces the r.h.s.  we further see that a
reduction of the reservoir of thermal energy entails a further
reduction in the growth in entropy due to particle production.
 
Since entropy increase was rather small anyway, it is now possible to
suppose that we have $S/S_0\simeq 1$. Then we find from
Eq.\,(\ref{flow2}) that the 
temperature cooling on the way to Hagedorn equilibrium is given by:
\begin{eqnarray}
{T_{\rm H}\over T_0}\simeq {1\over{1+\alpha_0/4}}
     \left(1-{E_{\rm H}\over E_0}\right)\, ,\label{flow3}
\end{eqnarray}
where $T_{\rm H}$ is the temperature at which the  Hagedorn
equilibrium is established, and there the fraction $E_{\rm H}/E_0$ of
energy was transferred to the flow. For an initial $\gamma_0=0.1$
occupancy, we have $\alpha_0=2.3$ and hence the cooling due to
particle production is by factor 0.63 (correct numerical Bose gas
result which accounts for entropy production is 0.55). The time period
is of the order of 1 fm/c and during this time the fraction of energy
transferred to flow may be as large as 40\%. Consequently, we see that
the particle production cooling (first factor in Eq.\,(\ref{flow3}))
is of the same importance as the hydrodynamical expansion cooling
(second factor in Eq.\,(\ref{flow3})) and that both combine to
increase the temperature reduction, while both do not lead to
appreciable entropy increase. We see that the cooling shown in
Fig.\,\ref{F2} is certainly of equal if not greater importance than
the cooling expected from the longitudinal hydrodynamic expansion of
the glue-parton volume obtained in a recent numerical study
\cite{Bir93}.
 
In summary, reviewing an old topic we have found that contrary to
intuitive expectations the entropy growth is not proportional to the
particle number increase, and it is indeed small during approach to
Hagedorn equilibrium ($\gamma\to 1$) by a hot glue gas. We have used
this result to show that the temperature cooling due to particle
production is comparable to cooling due to hydrodynamic expansion. We
have seen in detail why a sizable increase in gluon number is not
generating a large increase in entropy in the nuclear collision.  Our
finding thus renders the mechanism of entropy production in high
energy nuclear collisions even less understandable, as it now must
occur in the initial (quantum) time period before 0.3 fm/c 
\cite{Gei92b}, when the primordial hot glue phase is generated. 
  
{\it Acknowledgments:} We thank U. Heinz for a careful reading of the
manuscript and helpful comments.


\begin{thebibliography}{9}

\bibitem[$\dagger$]{JR}Work supported by DOE, grant DE-FG02-92ER40733  
 
\bibitem{Let93}
J. Letessier, A. Tounsi, U. Heinz, J. Sollfrank and J. Rafelski, {\it
Phys. Rev. Lett.} {\bf 70}, 3530 (1993).
 
\bibitem{Shu92}
E.V. Shuryak, {\it Phys. Rev. Lett.} {\bf 68}, 3270 (1992).
 
\bibitem{Gei92b}
K. Geiger {\it Phys. Rev.} {\bf D46}, 4986 (1992).
 
\bibitem{Bir93}
T.S. Bir\'{o}, E. van Doorn, B. M\"uller, M.H. Thomas and X.N.
Wang, {\it  Phys. Rev.} {\bf C48}, 1275 (1993).
 
\bibitem{Hag73}
R. Hagedorn, {\it Carg\`ese lectures in Physics}, Vol. {\bf 6}, Gordon
and Breach, N.Y. 1973.
 
\bibitem{deG80}
S.R. de Groot, W.A. van Leeuwen and Ch.G. van Weert, {\it
Relativistic Kinetic Theory} North Holland Pub. Co. Amsterdam 1980.   
 
\bibitem{OH93}
S. Ochs and U. Heinz, {\it Entropy Production by Resonance Decay and in
the $\rho$--$\pi\pi$ system}, manuscript in preparation, Regensburg
1993 and: W. Ochs, Diploma Thesis, Universit\"at Regensburg 1993.
 
\bibitem{Lan?}
L. Landau and E. Lifshitz, {\it Statistical Physics}, Sections 40, 54,
Pergamon Press, London 1959.
 
\bibitem{SH92}
J. Sollfrank and U. Heinz, {\it Physics Letters} {\bf B289}, 132
(1992).
 
\bibitem{Raf91}
J. Rafelski, {\it Phys. Lett.} {\bf B262}, 333 (1991) and {\it Nucl.
Phys.} {\bf A544}, 279c (1992).
 

\end{thebibliography}
\end{document}